\documentclass[aps,preprint]{revtex4}
\usepackage{amsmath}
\usepackage{blindtext}
\usepackage{extarrows}
\usepackage{braket}
\usepackage{pdfpages}
\usepackage{graphicx}
\usepackage{caption}
\usepackage{float}
\graphicspath{ {images/} }
\usepackage{amsmath,graphicx,color,xcolor,epsfig}
\usepackage{epstopdf}
\usepackage{hyperref}

\providecommand{\U}[1]{\protect\rule{.1in}{.1in}}

\begin{document}

\preprint{}
\title{Nonlinear Dynamics of Coherent Parametric Amplification in Multipartite two-level System under Intrinsic Decoherence}
\author{M. Ibrahim$^{1}$}
\email{ibrahim@phys.qau.edu.pk}

\affiliation{$^{1}$Department of Physics, Quaid-i-Azam University, Islamabad}

\date{\today }

\begin{abstract}
In this work, we study the dynamics of global quantum discord and quantum Fisher information in a multipartite system of two-level atoms interacting with a coherent field. The model includes parametric amplification, Kerr-type nonlinearity, and intrinsic decoherence to examine how these effects control quantum correlations and parameter-estimation sensitivity. The results show that, without intrinsic decoherence, both quantities exhibit rapid oscillations with clear collapse and revival behavior. Strong Kerr nonlinearity and strong parametric amplification enhance global quantum discord, while quantum Fisher information becomes maximum under a suitable balance of Kerr nonlinearity and amplification strength. Increasing the number of atoms generally strengthens global quantum discord but does not always improve quantum Fisher information. Intrinsic decoherence damps the oscillations and drives the system toward steady-state behavior.

\end{abstract}

\maketitle

\section{\qquad Introduction}

The field of quantum information processing (QIP) has witnessed an extraordinary transformation, driven by the unique properties of quantum correlations that lack classical counterparts. While entanglement has historically been regarded as the primary resource for quantum communication and computation \cite{1,2}, it is now widely recognized that entanglement does not encompass the totality of quantum correlations. This realization has led to the emergence of Quantum Discord (QD) as a more general measure of non-classicality, capable of identifying quantum advantages even in separable (non-entangled) states \cite{3,4,5}. In multipartite settings, where the complexity of interactions scales significantly, the study of Global Quantum Discord (GQD) becomes essential for characterizing the total quantum correlations distributed across a many-body system \cite{6,7}.

Simultaneously, the precision of quantum sensors and metrological tasks is fundamentally governed by Quantum Fisher Information (QFI). QFI serves as the central metric in quantum estimation theory, providing a lower bound on the variance of an estimated parameter through the quantum Cramér-Rao inequality \cite{8,9,10,11}. Specifically, for a multipartite system, the QFI determines the sensitivity of the state with respect to unitary transformations, where a value exceeding the "shot-noise limit" indicates the presence of useful multipartite correlations \cite{12, 13, 14}. Understanding the interplay between GQD and QFI is therefore paramount for the development of robust quantum technologies that rely on high-precision measurement and long-lived correlations.

The physical realization of these quantum resources often involves two-level atomic systems interacting with electromagnetic fields. To enhance and manipulate these correlations, non-linear optical elements such as the Kerr medium and parametric amplification are frequently employed. The Kerr medium, characterized by a third-order non-linear susceptibility, is known to generate non-classical states and protect quantum coherence against environmental noise \cite{15, 16, 17, 18}. Parametric amplification, on the other hand, allows for the generation of squeezed states, which are vital for suppressing quantum fluctuations and significantly boosting the QFI of a system \cite{20, 21, 22}. The synergy between these non-linearities offers a promising avenue for controlling the dynamics of multipartite atomic systems.

However, any realistic quantum system is inevitably subject to decoherence, which degrades quantum resources over time. In addition to external reservoir-induced decoherence, the intrinsic decoherence model proposed by Milburn \cite{19} provides a fundamental framework for studying the decay of quantum coherence. Unlike standard Markovian master equations, the Milburn model assumes that the system does not evolve continuously under the Schrödinger equation but rather through a stochastic sequence of unitary steps, leading to a natural phase-damping effect. Investigating how these intrinsic mechanisms influence a multipartite system driven by a coherent field is crucial for establishing the operational limits of quantum metrology and correlation distribution in practical environments.

Despite the extensive literature on bipartite correlations, the simultaneous evolution of GQD and QFI in multipartite systems under the combined influence of Kerr non-linearity, parametric amplification, and intrinsic decoherence remains an open question. Previous studies have examined the dynamics of entanglement and discord in two-qubit systems under various noise models \cite{24, 25}. However, a comprehensive analysis that bridges these non-linear optical effects with the Milburn decoherence model in a multi-atom context is currently lacking.

In this work, we investigate the joint dynamics of GQD and QFI for a multipartite two-level atomic system interacting with a coherent field, specifically accounting for the effects of a Kerr medium and parametric amplification. By solving the Milburn master equation, we derive the time-evolved density matrix and quantify the degradation of quantum correlations and metrological sensitivity. Our results highlight the critical roles played by the Kerr non-linearity and amplification parameters in mitigating the effects of intrinsic decoherence, offering insights into the preservation of quantum resources for advanced metrological applications.\\

\section{Hamiltonian Model}

In this work, we study an extended form of the Tavis-Cummings model \cite{tavis}, which provides a basic theoretical framework for describing multipartite quantum systems. In its original form, the model describes two identical two-level atoms, denoted by A and B, coupled to a single-mode quantized radiation field. Here, we generalize this model by introducing Kerr-type nonlinearity and parametric amplification, and we analyze the cases of two, three, and four atoms placed inside a cavity.

The Kerr effect originates from the third-order nonlinear response of the medium, which produces an intensity-dependent phase variation in the field. The refractive index in the presence of Kerr nonlinearity can be written as \cite{kerr1, kerr2, kerr3}
\begin{equation}
	n = n_0 + n_2 E^2,
\end{equation}
where $n_0$ represents the linear refractive index, while $n_2$ denotes the Kerr coefficient that measures the strength of the nonlinear response.

Along with Kerr nonlinearity, the present model also includes degenerate parametric amplification. This process is a second-order nonlinear optical effect in which one pump photon of frequency $2\omega$ is converted into two photons of frequency $\omega$. Such a process requires a nonzero second-order susceptibility, $\chi^{(2)}$, and can be produced by applying a classical pump field to a nonlinear medium. In the Schrödinger picture, the effective Hamiltonian corresponding to the parametric interaction is given by \cite{milburnnbook}
\begin{equation}
	H_{PA} = -i\hbar \frac{\kappa}{2} \left(a^2 e^{2i\omega t} - a^{\dagger 2} e^{-2i\omega t} \right),
\end{equation}
where $a$ and $a^\dagger$ denote the annihilation and creation operators of the field mode, respectively, and $\kappa$ represents the parametric pumping amplitude, which depends on the pump strength and the nonlinear susceptibility.

By applying the rotating wave approximation, the time-dependent factors are removed, and the Hamiltonian in the interaction picture becomes
\begin{equation}
	H_{PA} = -i\hbar \frac{\kappa}{2} \left(a^2 - a^{\dagger 2} \right).
\end{equation}

After including all the relevant interactions, the total Hamiltonian for a system of $N$ two-level atoms is expressed as
\begin{equation} \hat{H}_T = \frac{\omega_0}{2} \sum_{i=1}^{N} \hat{\sigma}_i^z + \omega \hat{a}^\dagger \hat{a} + g \sum_{i=1}^{N} \left( \hat{a} \hat{\sigma}_i^+ + \hat{a}^\dagger \hat{\sigma}_i^- \right) + \chi \left( \hat{a}^\dagger \hat{a} \right)^2 - i\hbar \frac{\kappa}{2} \left( a^2 - a^{\dagger 2} \right), \end{equation}
where $\omega_0$ and $\omega$ denote the atomic transition frequency and the field frequency, respectively. The operators $\hat{\sigma}_i^z$ and $\hat{\sigma}_i^\pm$ are the inversion and raising/lowering operators of the $i$-th atom, while $g$ represents the atom-field coupling strength. The term containing $\chi$ describes the Kerr-type nonlinear contribution.

The initial state of the total system is taken as the direct product of a partially mixed atomic state and a coherent field state:
\begin{equation}
	\hat{\rho}(0) = \left[ (1-p)|\psi\rangle\langle\psi| + p |g_1 g_2 \ldots g_N\rangle\langle g_1 g_2 \ldots g_N| \right] \otimes \hat{\rho}_E,
\end{equation}
where $0 \leq p \leq 1$ describes the degree of mixedness. The pure atomic state $|\psi\rangle$ is written as
\begin{equation}
	|\psi\rangle = \cos(\theta) |g_1 g_2 \ldots g_N\rangle + \sin(\theta) |e_1 e_2 \ldots e_N\rangle,
\end{equation}
with $0 \leq \theta \leq \pi$, and $\hat{\rho}_E$ represents the coherent field state $|\alpha\rangle$:
\begin{equation}
	\hat{\rho}_E = |\alpha\rangle\langle\alpha|.
\end{equation}

The time-dependent state of the atom-field system is given by
\begin{equation} \hat{\rho}_{AF}(t) = \sum_{i,j} \langle \psi_i | \hat{\rho}(0) | \psi_j \rangle e^{-i(E_i - E_j)t} |\psi_i\rangle \langle \psi_j|. \end{equation}
The reduced density matrix of the atomic subsystem is obtained by taking the trace over the field variables:
\begin{equation} \hat{\rho}_T(t) = \mathrm{Tr}_F \left[ \hat{\rho}_{AF}(t) \right]. \end{equation}
\section{Multipartite Quantum Correlations and Quantum Fisher Information}

In the early stage, a large part of research in quantum information theory was mainly focused on the study of entanglement in bipartite systems. For a composite quantum system formed by two subsystems, $A$ and $B$, quantum discord $D^{A \rightarrow B}$ measures the non-classical correlations that may still exist even in the absence of entanglement. It is defined as the difference between the quantum mutual information $I(\rho)$ and the classical correlation $J(\rho)$, where the classical part is minimized over all orthogonal projective measurements ${\hat{\varPi}}$ performed on subsystem $B$:
\begin{equation} D^{A\rightarrow B}(\rho_{AB}) = \min_{\{\hat{\varPi}_B^j\}} [I(\rho_{AB}) - J(\rho_{AB})_{\{\hat{\varPi}_B^j\}}]. \end{equation}

This idea has been generalized to describe correlations in multipartite quantum systems by introducing global quantum discord (GQD). For an $N$-partite system, the GQD is expressed as:
\begin{equation}
	GQD(\rho_T) = \min_{{\hat{\varPi}^j}} \left[ S(\rho_T | \hat{\varPi}(\rho_T)) - \sum_{j=1}^N S(\rho_j | \hat{\varPi}_j(\rho_j)) \right],
\end{equation}
where $\rho_T$ represents the total state, $\rho_j$ is the reduced density matrix of the $j$th subsystem, and $S(\rho_1 | \rho_2)$ denotes the relative entropy between two quantum states. This extension gives a broader way to quantify quantum correlations shared among all parts of a multipartite system.

For easier numerical calculation, another useful form of the GQD has been proposed \cite{gqd}:
\begin{equation} 
	GQD(\rho_T) = \min_{\{\varPi^k\}} \left\{ \sum_{j=1}^{N} \sum_{l=0}^{1} \tilde{\rho}_j^{ll} \log_2 \tilde{\rho}_j^{ll} - \sum_{k=0}^{2^N - 1} \tilde{\rho}_T^{kk} \log_2 \tilde{\rho}_T^{kk} \right\} + \sum_{j=1}^N S(\rho_j) - S(\rho_T), 
\end{equation}
where $\tilde{\rho}_T^{kk} = \langle k | \hat{R}^\dagger \rho_T \hat{R} | k \rangle$ and $\tilde{\rho}_j^{ll} = \langle l | \hat{R}^\dagger \rho_j \hat{R} | l \rangle$. Here, $\hat{\varPi}^k = \hat{R} |k\rangle \langle k| \hat{R}^\dagger$ are the projective operators, and $\hat{R}$ is a rotation operator defined as $\hat{R} = \bigotimes_{j=1}^N \hat{R}_j(\theta_j, \phi_j)$, where $\hat{R}_j(\theta_j, \phi_j) = \cos\theta_j \hat{1} + i \sin\theta_j \cos\phi_j \hat{\sigma}_y + i \sin\theta_j \sin\phi_j \hat{\sigma}_x$.

In quantum metrology, quantum Fisher information (QFI) is an important quantity for estimating an unknown parameter $\theta$ with high accuracy. The classical Fisher information (CFI) is written as:
\begin{equation}
	I_\Phi = \sum_i p_i(\theta) \left( \frac{\partial}{\partial\theta} \ln p_i(\theta) \right)^2,
\end{equation}
where $p_i(\theta)$ denotes the probability distribution of the measurement outcomes associated with the parameter $\theta$.

QFI generalizes this concept to quantum systems and provides the fundamental precision bound for parameter estimation. It is defined as:
\begin{equation}
	F_\Phi = \text{Tr}[\rho(\theta) D^2],
\end{equation}
where $D$ is the symmetric logarithmic derivative (SLD), determined by:
\begin{equation}
	\frac{d\rho(\theta)}{d\theta} = \frac{1}{2} \left[ \rho(\theta) D + D \rho(\theta) \right].
\end{equation}

The spectral representation of the density matrix $\rho_\theta$ is given by:
\begin{equation} \rho(\theta) = \sum_k \lambda_k |k\rangle \langle k|, \end{equation}
with QFI given by:
\begin{equation}
	F_\theta = \sum_k \frac{(\partial_\theta \lambda_k)^2}{\lambda_k} + 2 \sum_{k,k'} \frac{(\lambda_k - \lambda_{k'})^2}{\lambda_k + \lambda_{k'}} |\langle k | \partial_\theta k' \rangle|^2,
\end{equation}
where $\lambda_k > 0$ and $\lambda_k + \lambda_{k'} > 0$. The first part corresponds to the classical Fisher information, whereas the second part represents the purely quantum contribution.

For the evaluation of the average QFI (AQFI) in a composite system, the field degrees of freedom are traced out. For a bipartite density matrix $\rho_{AB}$, AQFI is given as:
\begin{equation}
	I_{QF}(t) = \text{Tr}[\rho_{AB}(\theta, t) D(\theta, t)^2],
\end{equation}
with the corresponding SLD $D(\theta, t)$ satisfying:
\begin{equation}
	\frac{\partial \rho_{AB}(\theta, t)}{\partial \theta} = \frac{1}{2} \left[ D(\theta, t) \rho_{AB}(\theta, t) + \rho_{AB}(\theta, t) D(\theta, t) \right].
\end{equation}

\section{Numerical \textbf{results and discussions}}
We solve our multipartite system numerically and take the scaled time step size of 0.05. Since the higher order systems require significant amount of computational power, we limit our numerical calculations up to four two-level system. For the case of the QFI, we estimate the system parameter $\theta$ around the value $3\pi/4$. 
\subsection{The GQD and QFI dynamics for varying nonlinear Kerr medium parameter}
\begin{figure}[H]
	\centering
	\includegraphics[width=5.5in]{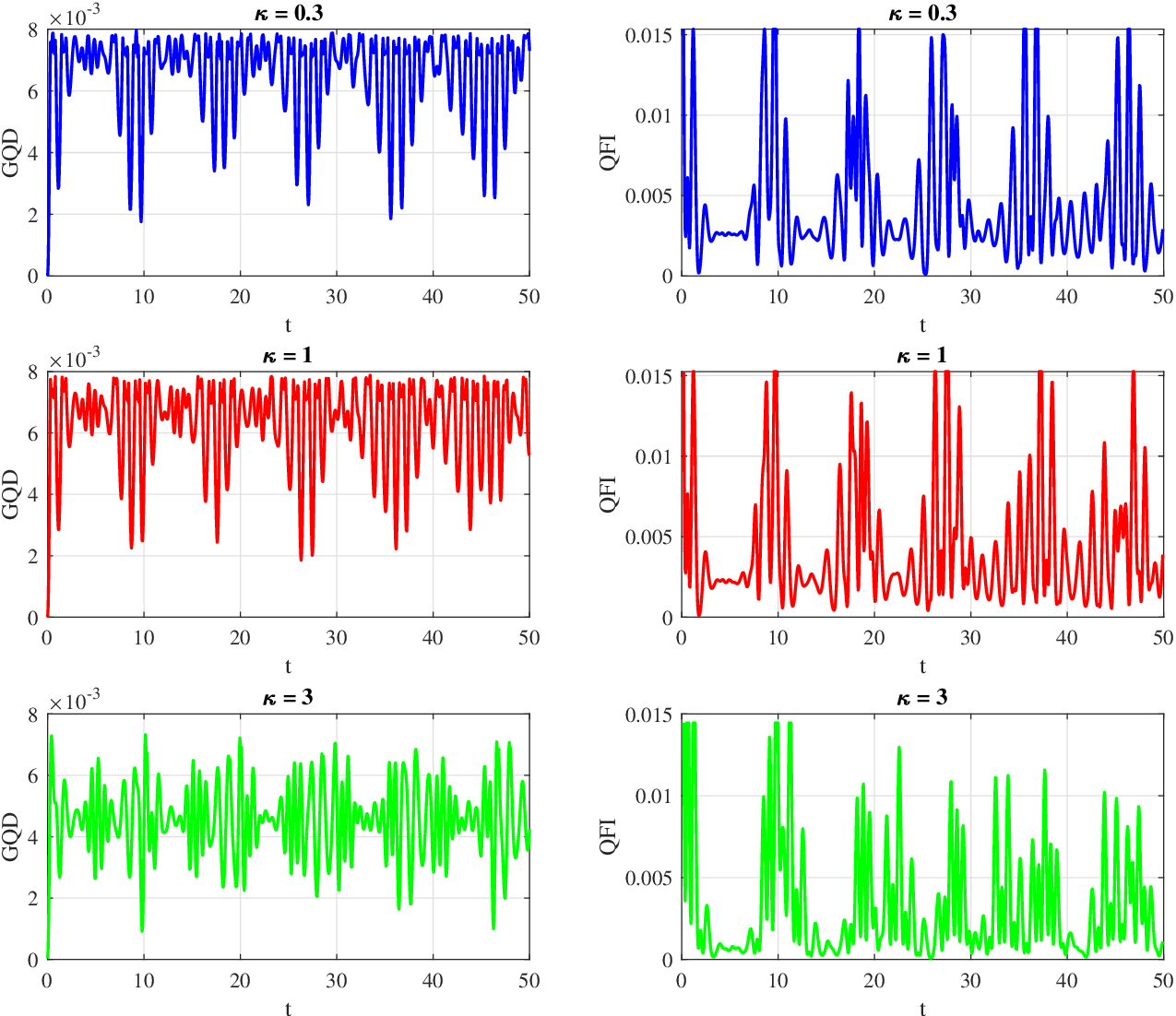}
	\caption{(color online) Dynamics of the GQD (left column) and QFI (right column) for a multipartite atomic system (N=2) coupled to a coherent field. The Kerr nonlinearity is fixed at $\chi=0.3$, the intrinsic-decoherence rate is set to $\gamma=0$, and the initial coherent-field amplitude is $|\alpha|=2$. From top to bottom, the parametric-amplification strength is varied as $\kappa=0.3$, $1$, and $3$, respectively. The results show that moderate parametric amplification ($\kappa=1$) enhances the quantum correlations and improves the parameter-estimation sensitivity. In contrast, strong parametric amplification ($\kappa=3$) introduces pronounced nonlinear fluctuations, leading to the suppression of long-time coherence, global quantum correlations, and metrological sensitivity.}
	\label{fig1}
\end{figure}

\begin{figure}[H]
	\centering
	\includegraphics[width=5.5in]{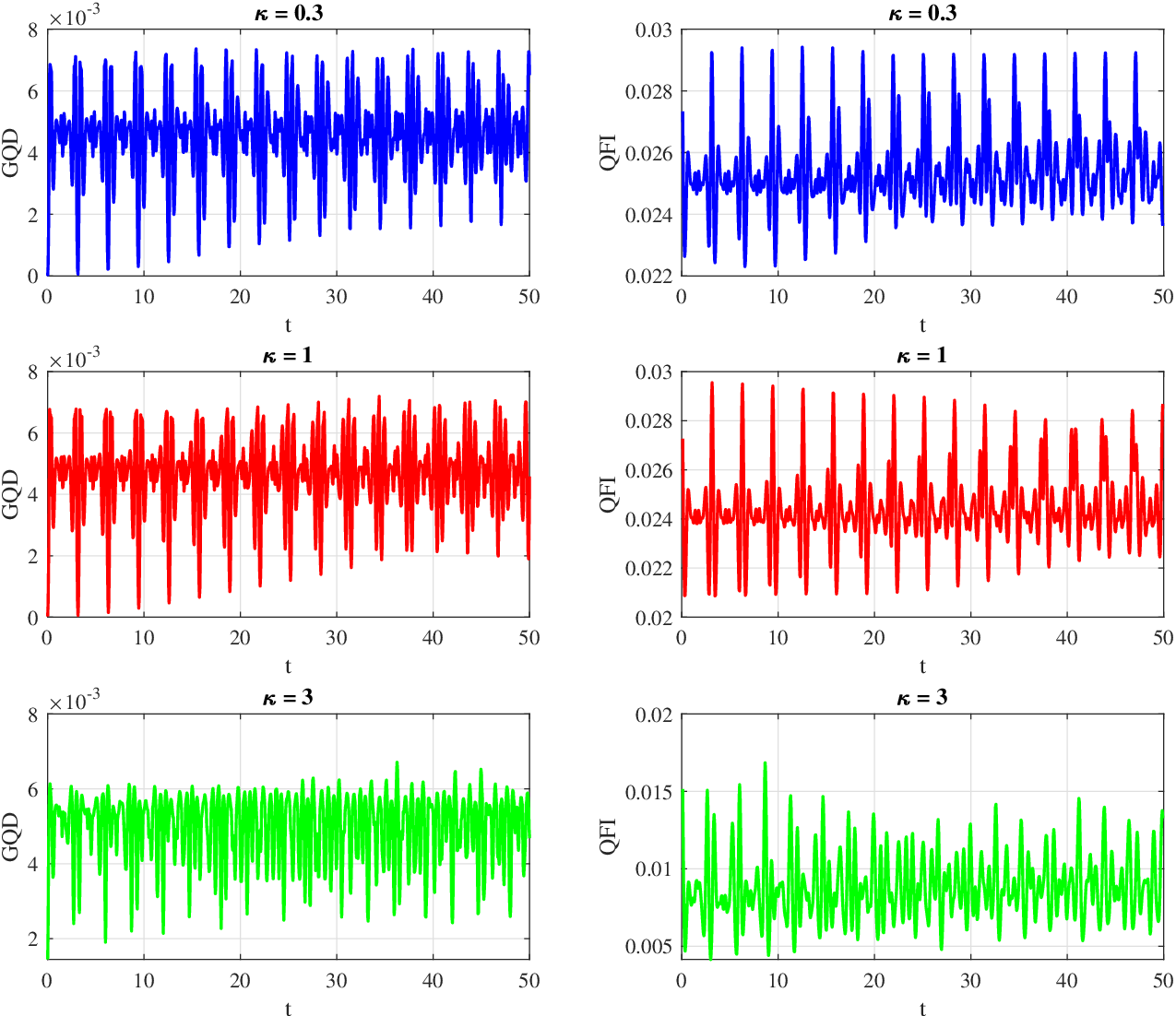}
	\caption{(color online) Dynamics of the GQD and QFI for nonlinear Kerr parameter $\chi=1$, with all remaining parameters identical to those in Fig.(\ref{fig1}). In this relative stronger Kerr regime, large values of $\kappa$ enhance nonlinear phase dispersion and induce effective dephasing, thereby suppressing quantum coherence. As a result, both quantum correlations and Fisher information are reduced.}
	\label{fig2}
\end{figure}

\begin{figure}[H]
	\centering
	\includegraphics[width=5.5in]{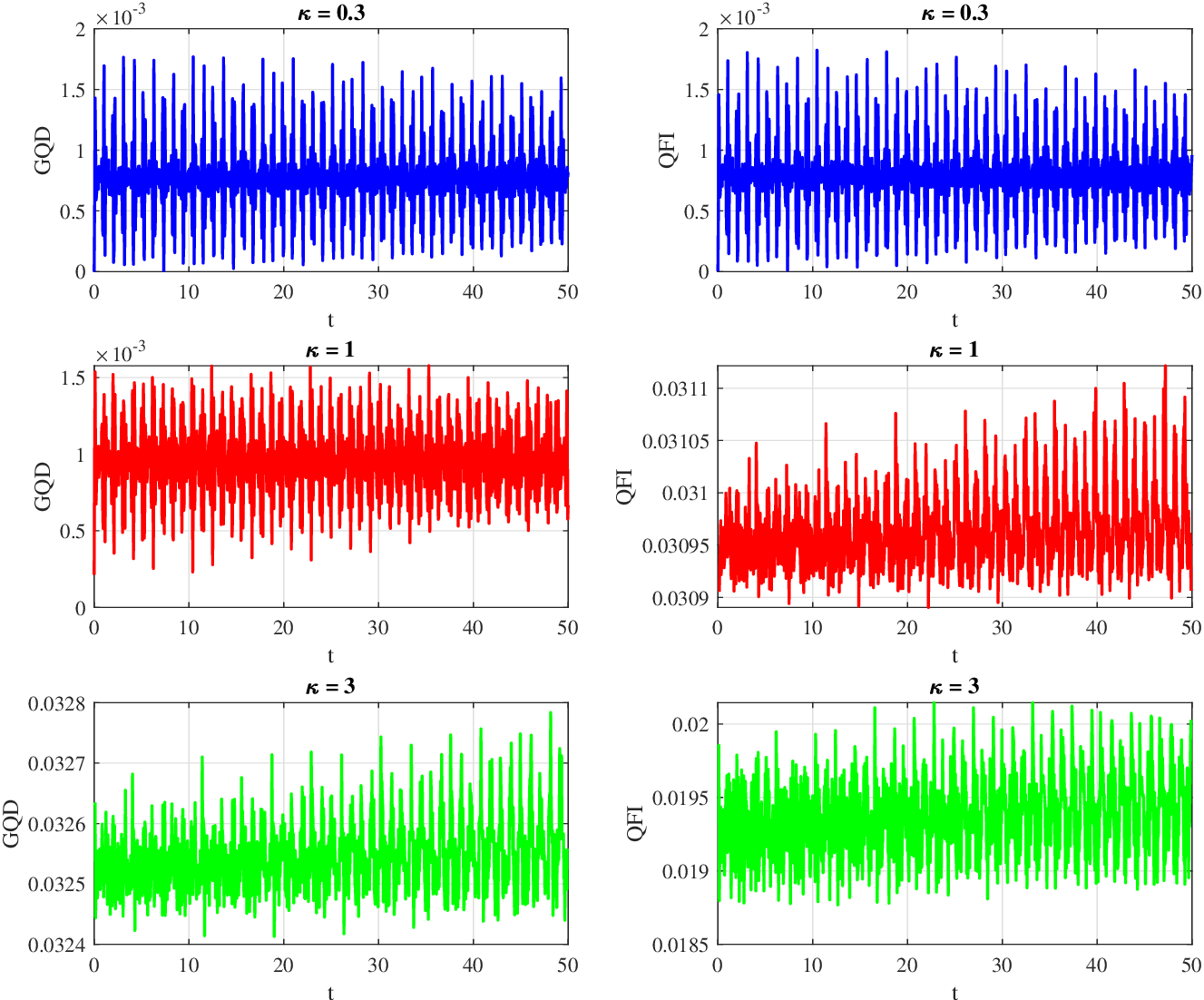}
	\caption{(color online) Dynamics of the GQD and QFI for nonlinear Kerr parameter $\chi=3$, with all remaining parameters identical to those in Fig. (\ref{fig1}). In this strong-Kerr regime, larger $\kappa$ values are more favorable for sustaining and enhancing the GQD. However, the maximum QFI, and hence the optimal parameter-estimation precision, is obtained at the intermediate amplification strength $\kappa=1$.}
	\label{fig3}
\end{figure}

In this section we study the dynamics of the GQD and QFI of our system. For each figure we change the nonlinear Kerr constant $\chi$ and in a figure we change the $\kappa$ down the row. We set $\gamma=0$ for this section. We have taken the average coherent field photons $|\alpha|=2$.\\
Fig. (\ref{fig1}) shows the dynamics of the GQD and QFI for a two two-level atomic system for different strengths of parametric amplification coefficient $\kappa$. The Kerr constant is fixed at $\chi=0.3$. The value of $\kappa$ is increases from $0.3$ to $3$ for the first row to the last row. The left and right columns represent the GQD and QFI, respectively. For $\kappa=0.3$, the GQD exhibits oscillatory behavior of the dynamics. Although a sudden periodic suppression appears at certain time intervals, the GQD rapidly revives. The corresponding QFI dynamics for $\kappa=0.3$ display intermittent peaks reaching $1.5\times 10^{-2}$. The GQD and QFI show revival peaks at the same time interval, indicating that the quantum correlations enhance with improved parameter estimation. The average value of the GQD remains at around $8\times 10^{-3}$ and QFI reside just below around average value of $0.005$ in the dynamics. The QFI average value fluctuate more irregularly and its value increases over time. The peaks in the QFI indicate temporary enhancement of phase sensitivity of $\alpha$ and its parameter estimation precision. Between the peaks, the GQD and QFI remain finite, showing that the quantum state retains useful quantum correlations and metrological information throughout the evolution. When the amplification coefficient increases to $\kappa=1$, the GQD and QFI dynamics exhibit similar behavior as for $\kappa=0.3$ case. The QFI oscillations become more irregular with the time evolution. The packet like revivals of the QFI also become fade as time progress. The average value of the GQD and QFI remains the same as for the case of $\kappa=0.3$. For the strong parametric amplification regime $\kappa=3$, the GQD dynamics change significantly. The average value decreases as compared to the previous cases. The oscillations become irregular and are now nearly symmetric around the average value. The amplitude of oscillations is suppressed. The QFI dynamics for $\kappa=3$ is characterized by localized sharp peaks separated by intervals of small value. This suggest that the system experiences short lived episodes of high matrological sensitivity followed by period of reduced quantum coherence. Overall, Fig. (\ref{fig1}) demonstrate that parametric amplification coefficient $\kappa$ plays a crucial role in the system dynamics. Moderate amplification enhances quantum correlations and parameter estimation sensitivity. Whereas, large amplification introduces nonlinear fluctuations that suppress long term coherence.\\
Fig. (\ref{fig2}) represents the dynamical evolution of the GQD and QFI for a different value of nonlinear Kerr medium parameter, $\chi=1$. The system under study is a two two-level atomic system (N=2). The dynamics is studies for different value of parametric amplification strengths. For $\kappa=0.3$, the GQD oscillates around an average value of nearly $5\time 10^{-3}$. The system observe revival like oscillation and revivals are accompanied by sharp periodic suppression approaching zero in early dynamics. The dynamics indicate repeated collapses and revivals of quantum correlation. The corresponding QFI oscillates around $2.5\times 10^{-2}$ with regular peaks close to $3\times 10^{-2}$. Both GQD and QFI exhibit unsymmetrical amplitude of oscillations around its average value. At $\kappa=1$, the oscillations in the GQD become more denser. Overall the dynamics of both GQD and QFI remain the same for this intermediate amplification strength with minor modification. The GQD amplitude decreases as time progress. On the other hand, the QFI average value decreases in the dynamics. This indicates that as time progresses the quantum correlations of the system are decreased with decreased parameter estimation. For strong amplification strength, $\kappa=3$, the GQD fluctuations become highly irregular with reduced depth of revivals. The amplitude of revivals is also decreased, oscillating mainly between $3\times 10^{-3}$ and $6\times 10^{-3}$. In contrast, the QFI decreases considerably remaining mostly below $1.5\times 10^{-2}$. The large values of $\kappa$ introduces dephasing in quantum coherence that reduces both quantum correlations and metrological precision.
Fig. (\ref{fig3}) describes the affect of strong nonlinear Kerr medium parameter in the dynamics of the quantifiers for variable strength of parametric amplification strengths $\kappa$. For weak amplification case, $\kappa=0.3$, both GQD and QFI remain nearly similar in dynamical behavior. The average value of quantum correlations remains just below $1\times 10^{-3}$, and the QFI also lies in this range. It is seen that both GQD and QFI amplitudes decreases as time progresses. The GQD fluctuates rapidly around approximately $0.7\times 10^{-3}$ to $0.9\times 10^{-3}$ with sharp peaks reaching nearly close to $1.7\times 10^{-3}$ and with occasional minima close to zero. When $\kappa$ is increased to 1, the behavior of both quantifiers becomes different. The GQD still remains in $\times 10^{-3}$ range, oscillating mostly around $0.8\times 10^{-3}$ to $1.1\times 10^{-3}$ with peaks close to $1.5\times 10^{-3}$. We observe no collapse close to zero for this case and the revival amplitude decreases as time progresses. However, the QFI increases very strongly, and reaches values around $0.0309$ to $0.0311$. This is much larger than the magnitudes of the QFI at $\kappa=0.3$. Therefore intermediate amplification strength ($\kappa=1$) greatly improves the parameter estimation, even though the global discord does not increase in the same proportion. Furthermore, the QFI amplitude increases as time progresses. For the strong amplification case, $\kappa=3$, the GQD amplitude is enhanced as compared to previous cases. Its value oscillates between $0.0324$ to $0.0328$. The oscillations in the GQD are still rapid, but the amplitude curve shifted upwards showing that strong parametric amplification produces a more correlated multipartite quantum state. The QFI for $\kappa=3$ remain significant, oscillating between $0.0188$ and $0.021$, but its value is reduced as compared to the intermediate case. This implies that strong amplification is more favorable for the GQD while the best estimation of parameter appears at intermediate value $\kappa=1$.\\
Overall, for Figs. (\ref{fig1}-\ref{fig3}), show that the Kerr parameter ($\chi$) and the parametric amplification coefficient ($\kappa$) affect the quantifiers in different ways. For small Kerr nonlinearity, $\chi=0.3$, the system exhibits large GQD of the order $\times 10^{-3}$ and intermittent QFI peaks especially for $\kappa=0.3$ and $\kappa=1$. When $\chi$ is increased to 1, we observe strongly enhanced QFI, gaining the values close to $0.03$ for weak and intermediate amplification. Whereas, the GQD remains in the $10^{-3}$ range. For strong Kerr nonlinearity, $\kappa=3$, the response depends very sensitively on parametric amplification coefficients $\kappa$. Weak amplification suppresses both quantifiers, intermediate amplification increases the QFI and strong amplification assists the GQD. Hence the condition to obtain maximum quantum correlations is not same the condition to have maximum parameter estimation. The enhanced GQD is obtained for the combined strong Kerr and amplification case ($\chi=3$, $\kappa=3$), while the maximum QFI appears for strong Kerr but intermediate parametric amplification case ($\chi=3$, $\kappa=1$).
\subsection{Effect of increasing parametric amplification coefficient and atomic system size on the dynamics of the GQD and QFI}
\begin{figure}[H]
	\centering
	\includegraphics[width=5.5in]{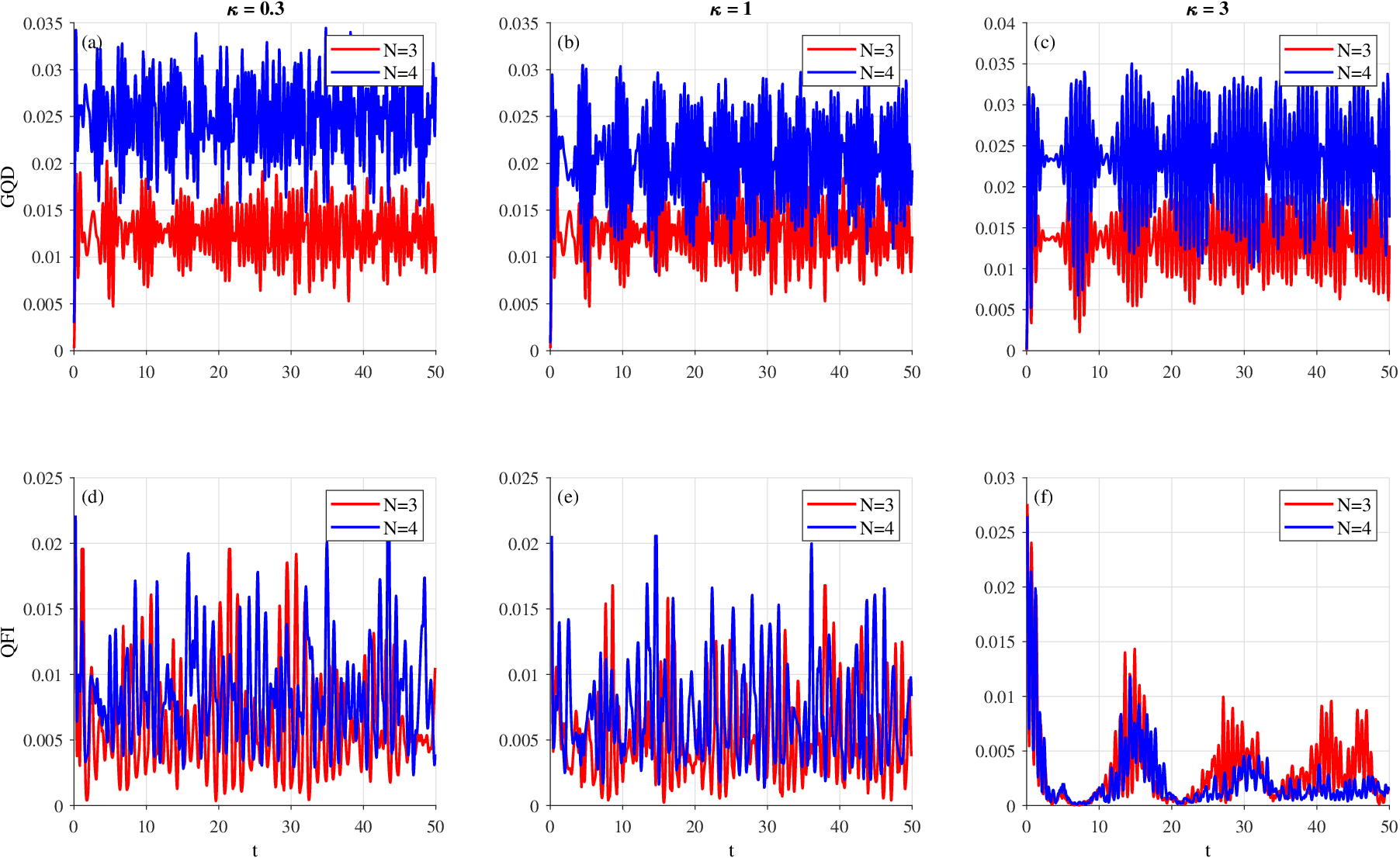}
	\caption{(color online) Dynamics of the GQD and QFI for multipartite systems composed of $N=3$ and $N=4$ two-level atoms under different parametric-amplification strengths $\kappa$, with nonlinear Kerr constant value $\chi=0.3$, $|\alpha|=2$ and $\gamma=0$. The GQD exhibits distinct revival-like packets for both system sizes, while its mean value is larger for $N=4$ than for $N=3$, demonstrating that an increased number of atoms enhances the overall quantum correlations. In contrast, for weak and intermediate parametric amplification, increasing the atomic-system size does not lead to an improvement in parameter-estimation sensitivity. Moreover, strong parametric amplification suppresses the long-time QFI, thereby reducing the achievable metrological precision.}
	\label{fig4}
\end{figure}

Fig. (\ref{fig4}) represents the dynamics of the GQD and QFI for two different multipartite systems composed of three ($N=3$) and four ($N=4$) two-level atomic systems under variable parametric amplification strengths. The nonlinear Kerr parameter value is fixed and taken $\chi=0.3$. For all value of $\kappa$, we observe that the GQD dynamics exhibit oscillatory behavior. The nature of these oscillations changes significantly as $\kappa$ increases. For $\kappa=0.3$, we observe rapid and irregular correlation oscillations around a finite mean value for both systems. The mean value for $N=4$ system is greater than $N=3$ system. The amplitude of $N=4$ system is also enhanced. For $\kappa=1$, the mean value of the GQD for $N=4$ system is decreased as compared to weak amplification case but the amplitude of revivals is enhanced. On the other hand, the mean value for $N=3$ system remains the same with nearly comparable amplitude relative to $\kappa=0.3$ case. The GQD mean value is enhanced as compared to $N=3$ case confirming that increasing the number of two-level atomic system increases the quantum correlations. The dynamical behavior is now expressing distinctive revival like packets for both systems. A more distinctive feature appears for the strong amplification case, $\kappa=3$, where multipartite quantum correlations display clear collapses and revival dynamics for both atomic configurations. The discord repeatedly decreases from a pronounced oscillatory correlation region (a revival) to a no oscillation interval (a collapse) of discord. For $N=3$, these collapses are relative deeper and the amplitude ranges from approximately $0.005$ to $0.018$. While for $N=4$, the mean value of discord is again high and the amplitude of revivals is also enhanced relative to $N=3$ case. For $N=4$ case, the amplitude of revivals ranges between $0.035$ to $0.01$. The collapses tend to fade as time progresses. The QFI dynamics show a different behavior from the GQD. At $\kappa=0.3$, both $N$ cases exhibit fast irregular oscillations describing the parameter estimation is highly sensitive in the dynamics. The amplitude of the QFI for both systems is nearly the same in the dynamics. For $\kappa=1$ the dynamics of the systems are again highly oscillatory and similar to $\kappa=0.3$ case. Thus for weak and intermediate amplification, increasing the number of atoms does not improve parameter estimation sensitivity. For $\kappa=3$, the QFI changes noticeably. Both $N=3$ and $N=4$ show a strong initial peaks near $t=0$ followed by a rapid decay to small oscillatory values. Both curves reach a large initial value around nearly $0.026$ and after a scaled time value $t=5$, both curves mostly remain below $0.010$ with revival peaks appearing at $t\approx 15$, $t\approx 30$ and $t\approx 43$. This implies that strong amplification suppresses the long-time QFI.\\ 
\subsection{Influence of Average Coherent Photons and Parametric Amplification on the GQD and QFI dynamics}
\begin{figure}[H]
	\centering
	\includegraphics[width=5.5in]{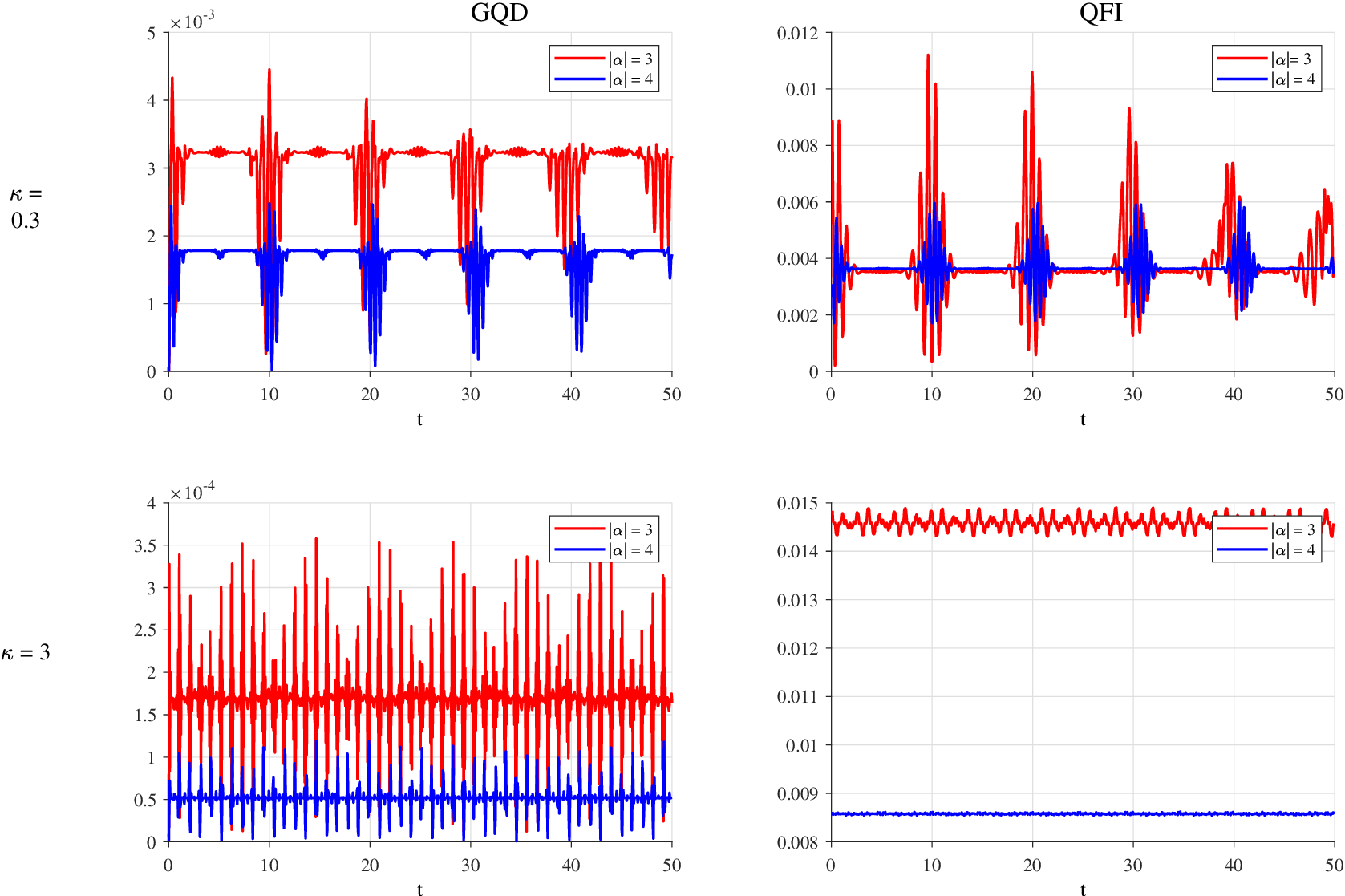}
	\caption{(color online) Dynamics of the GQD and QFI illustrating the effect of changing the coherent-field amplitude $|\alpha|$, with nonlinear Kerr medium parameter value $\chi=0.3$ and $\gamma=0$. Two parametric-amplification regimes are considered, namely $\kappa=0.3$ and $\kappa=3$. The results show that increasing the average number of coherent photons suppresses the multipartite quantum correlations, leading to lower GQD. For the strong-amplification case, the two photon-number settings yield different mean values of QFI, and these values remain enhanced compared with the weak-amplification regime, indicating improved parameter-estimation sensitivity despite the reduction in correlations.}
	\label{fig5}
\end{figure}

The effect of changing average photons of coherent field on the dynamics of the GQD and QFI for two different value of parametric amplification $\kappa$ is shown in Fig. (\ref{fig5}). For weak parametric amplification case, $\kappa=0.3$, the GQD exhibits a nearly periodic collapse-revival like structure. For $|\alpha|=3$, the GQD remains around $3.2\times 10^{-3}$, with sharp revival amplitude reaching approximately $4.4\times 10^{-3}$ and with minimum value falling below $1\times 10^{-3}$. When the average photons are increased to $|\alpha|=4$, the GQD is reduced considerably and stay mostly close to $1.7\times 10^{-3}$ with revival peaks around $2.4\times 10^{-3}$. The amplitude of revival is decreased with respect to $|\alpha|=3$ case. Furthermore the amplitude of revival of correlations reduces as time progresses for both average photon cases. This indicates that, in weak amplification regime, increasing mean photon number, suppresses the strength of quantum correlations. For weak amplification regime, the corresponding QFI shows a sequence of periodic revivals around $t\approx 0$, $10$, $20$, $30$, $40$ and $50$. For $|\alpha|=3$, these revival bursts are stronger, while the mean value of QFI remains the same, settling around just below $0.004$. For stronger parametric amplification, $\kappa=3$, the GQD is strongly suppressed and becomes of the order of $10^{-4}$. For $|\alpha|=3$, the GQD is further reduced and stays mostly around $0.5\times 10^{-4}$. Thus stronger parametric amplification weakens the GQD compared with weak amplification case. The increase in average photons in the system further suppresses the multipartite correlations. In contrast, the QFI behaves very differently at $\kappa=3$. Instead of collapse-revival bursts, it becomes almost steady over the whole time interval. For $|\alpha|=3$, the QFI remains nearly constant around $0.0145$ with only small oscillations, while for $|\alpha|=4$ it stays close to 0.0086. This implies that for strong amplification case, both photons cases have different average values of QFI and the values are enhanced as compared to weak case. This means that strong parametric amplification stablizes the QFI dynamics, but the increased average photons reduces the attainable parameter estimation. We conclude this section with that $|\alpha|=3$ case is more favorable that $|\alpha|=4$ for both GQD and QFI. Weak amplification supports clear collapse-revival behavior while strong amplification suppresses GQD but produces more stable QFI response.\\
\subsection{Effect of Atomic System Size and Parametric Amplification on GQD and QFI Dynamics under Intrinsic Decoherence}
\begin{figure}[H]
	\centering
	\includegraphics[width=5.5in]{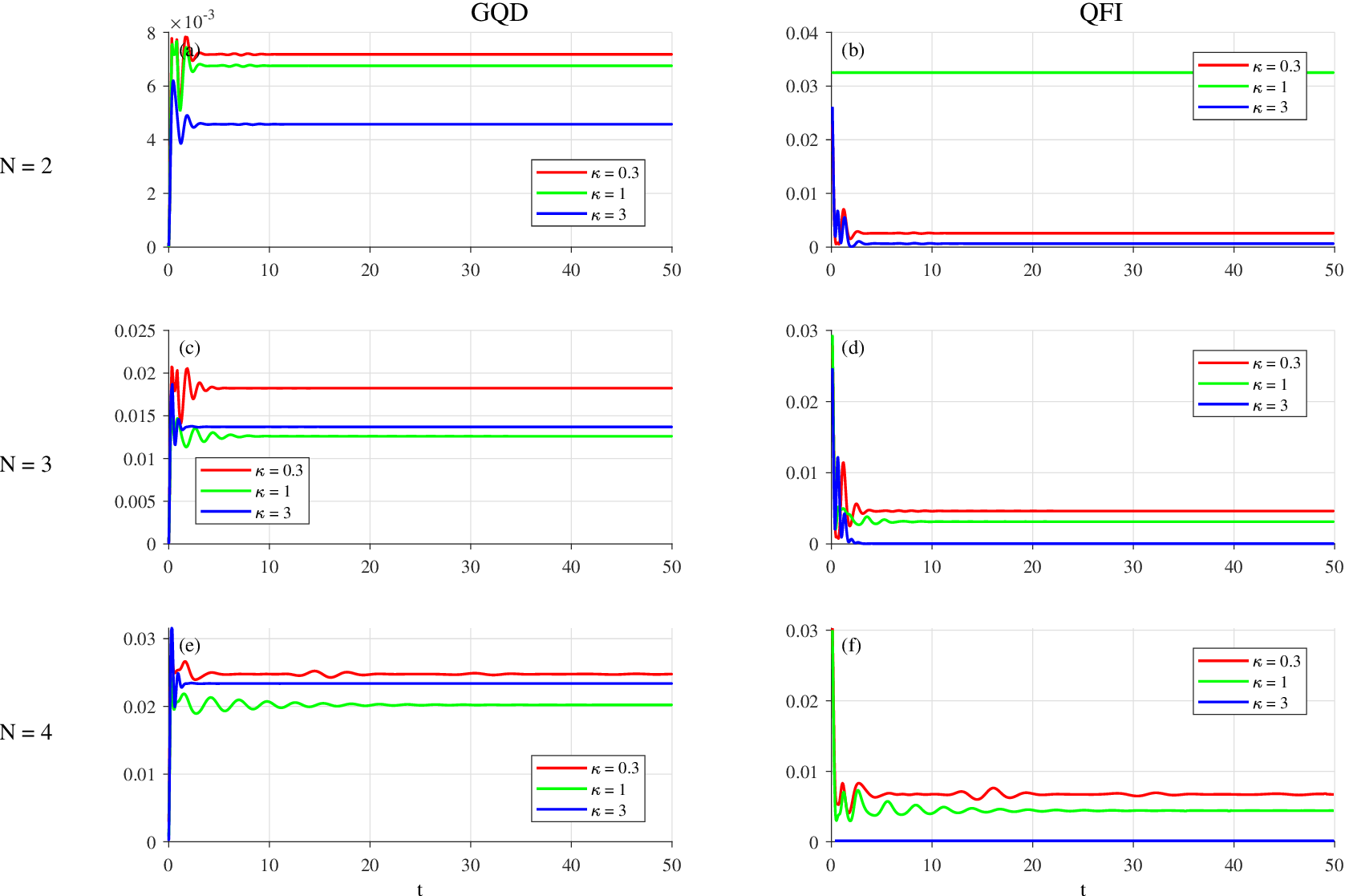}
	\caption{(color online) Dynamics of the GQD and QFI for multipartite systems with different numbers of two-level atoms, $N$, in the presence of intrinsic decoherence with $\gamma=0.01$. The remaining parameters are fixed at $\chi=0.3$ and $|\alpha|=2$. Intrinsic decoherence causes rapid damping of the oscillations in both GQD and QFI, demonstrating the progressive loss of multipartite quantum correlations and metrological sensitivity}
	\label{fig6}
\end{figure}

Fig. (\ref{fig6}) represents the time evolution of the GQD and QFI for different number of two two-level atomic system ($N$), under intrinsic decoherence of strength $\gamma=0.01$. We set $\chi=0.3$ and $|\alpha|=2$. The most visible effect of intrinsic decoherence is the rapid damping of oscillation in both quantum correlations and Fisher information. We observe that both GQD and QFI approaches steady state value approximately $t<5$. For $N=2$ system, the GQD initially shows a small damped oscillations and then settles near steady state value $7.1\times 10^{-3}$ for $\kappa=0.3$, around $6.8\times 10^{-3}$ for $\kappa=1$ and around $4.6\times 10^{-3}$ for $\kappa=3$. Thus for $N=2$, weak parametric amplification preserves the largest quantum correlation while strong amplification suppresses it. On the other hand, the QFI for this case behaves differently. For $\kappa=1$, the curve has a high average value close to $3.2\times 10^{-3}$ where $\kappa=3$ has least mean value settles near zero. For the case of $\kappa=0.3$, the QFI has near to zero average value compared to $\kappa=1$ around $0.002$. For $N=3$, the GQD increases considerably as compared to $N=2$ case. For $\kappa=0.3$, the curve reaches highest steady state value approximately $0.018$, while $\kappa=1$ and $\kappa=3$ settles near 0.015, with the GQD value at $\kappa=3$ increased compared to $\kappa=1$. The corresponding QFI is much smaller than the large ($N=2$, $\kappa=1$) case. The QFI value for $\kappa=0.3$ is greater than both other cases of $\kappa$, with steady state value at $\kappa=3$ nearly zero. Increasing the size $N$, affects only $\kappa=1$ value while other $\kappa$ values of QFI do not change much. For $N=4$, the GQD value further enhanced, confirming that increasing the number of atoms strengths multipartite quantum correlations, even in the presence of intrinsic decoherence. The steady state value of the GQD is $2.45\times 10^{-3}$ for $\kappa=0.3$, $2\times 10^{-3}$ for $\kappa=1$ and $2.3\times 10^{-3}$ for $\kappa=3$. The QFI for $N=4$ also reaches stable value after damped oscillations with largest value near $6.5\times 10^{-3}$ for $\kappa=0.3$ followed by $\approx 4.3\times 10^{-3}$ for $\kappa=1$ and again remains almost zero for $\kappa=3$. Overall, intrinsic decoherence removes the periodic collapse-revival behavior observed in decoherence free case and derive the system towards a steady state dynamics. Increasing $N$ strongly assists GQD with QFI is more sensitive to choice of $\kappa$. Weak amplification case ($\kappa=0.3$) is generally more favorable for sustaining the GQD for larger $N$. While the exceptional high QFI response occurs for $N=2$ at $\kappa=1$ case. \\

\section{Conclusions}
In this work investigated the dynamical behavior of the GQD and QFI of a multipartite two-level system interacting with coherent field. We introduced parametric amplification, non-linear Kerr effect, and intrinsic decoherence into the model and explored the effects of these parameters on the dynamics.\\
The results showed that the dynamics of the quantifiers, i.e. GQD and QFI were strongly dependent on the controlling parameters, i.e. parametric amplification coefficient ($\kappa$), the non-linear Kerr medium parameter ($\chi$), the size of the multipartite system ($N$), the mean photons in the coherent field ($|\alpha|$), and intrinsic decoherence ($\gamma$). In the absence of intrinsic decoherence, both quantifiers exhibited rapid oscillations with distinctive features of collapses and revivals. We observed that the GQD was enhanced when the system was subjected to strong Kerr effect and parametric amplification. On the other hand, maximum QFI was appeared for strong $\chi$ and weak $\kappa$ parameters. Increasing the number of atoms enhanced the GQD and increasing parametric amplification induced collapse-revival in the dynamics. Furthermore, for weak and intermediate amplification, increasing the system dimensions did not improved the QFI. When mean photons were increased, weak amplification caused collapses-revivals in both quantifiers. Strong amplification suppressed GQD for both cases of average photons, on the other hand it stabilized QFI response. When intrinsic decoherence was present, the steady state value of the GQD was increased for larger $N$. In the presence of intrinsic decoherence, the high QFI response was observed for $N=2$ case with intermediate value of parametric amplification.\\ 
The present study can be extended in several useful research directions. We can extend this model by taking different kind of quantum fields like thermal field, add detuning effects, considering atomic motion, and take different initial atomic states. It would be also interesting to study other entanglement measures such as von Neumann entropy, entanglement negativity, quantum coherence and geometric phase. Moreover, the optimization of $\kappa$, $\chi$, $\alpha$, and $N$ may provide directions to develop efficient quantum sensors and multipartite quantum information protocol. \\

\end{document}